\documentclass[12pt,preprint]{aastex}
\received{2005 August 30}
\begin{document}

\title{Beryllium in Disk and Halo Stars -- \\
Evidence for a Beryllium Dispersion in Old Stars\altaffilmark{1} }

\author{Ann Merchant Boesgaard\altaffilmark{2}} 

\and

\author{Megan C. Novicki\altaffilmark{2}} 

\affil{Institute for Astronomy, University of Hawai`i at M\-anoa, \\ 
2680 Woodlawn Drive, Honolulu, HI {\ \ }96822 \\ } 
\email{boes@ifa.hawaii.edu}
\email{mnovicki@ifa.hawaii.edu}

\altaffiltext{1}{Based on observations obtained with the Subaru Telescope}

\altaffiltext{2}{Open time observers at JNLT, Subaru Observatory,  operated
by the National Astronomical Observatory of Japan.}

\begin{abstract}
The study of Be in stars of differing metal content can elucidate the
formation mechanisms and the Galactic chemical evolution of the light element,
Be.  We have obtained high-resolution, high signal-to-noise spectra of the
resonance lines of Be II in eight stars with the high-dispersion spectrograph
(HDS) on the Subaru 8.2 m telescope on Mauna Kea.  Abundances of Be have been
determined through spectrum synthesis.  The stars with [Fe/H] values $>-$1.1
conform to the published general trend of Be vs.~Fe.  We have confirmed the
high Be abundance in HD 94028 and have found a similarly high Be abundance in
another star, HD 132475, at the same metallicity: [Fe/H] = $-$1.5.  These two
stars are 0.5 - 0.6 dex higher in Be than the Be-Fe trend.  While that general
trend contains the evidence for a Galaxy-wide enrichment in Be and Fe, the
higher-than-predicted Be abundances in those two stars shows that there are
also local Be enrichments.  Possible enrichment mechanisms include hypernovae
and multiple supernova explosions contained in a superbubble.  One of our
stars, G 64-37, has a very low metallicity of [Fe/] = $-$3.2; we have
determined its Be abundance to look for evidence of a Be plateau.  It's Be
abundance appears to extend the Be-Fe trend to lower Fe abundances without any
evidence for a plateau as had been indicated by a high Be abundance in another
very metal-poor star, G 64-12.  Although these two stars have similar Be
abundances within the errors, it could be that their different Be values are
indicators of may be indicating that a Be dispersion even at the lowest
metallicities.
\end{abstract}


\section{Introduction}

The study of Be in stars has importance in several areas of astronomy.  The
history of the production of Galactic Be can be seen in the increase of Be
abundance with stellar metallicity, usually measured by Fe or O
(e.g. Boesgaard et al.~1999) but other elements such as Ca and Mg have been
employed (e.g. King 2002).  The Fe abundance is used as a surrogate for age,
but the O abundance is relevant because it it directly connected to the
dominant, and perhaps only, Be production mechanism: spallation reactions in
the interstellar medium or in the vicinity of supernovae explosions.

The idea of spallation was invented by Reeves, Fowler \& Hoyle (1970) and the
details were first described by Meneguzzi, Audouze \& Reeves (1971).  The
basic process is that high energy ($\sim$150 MeV) protons and neutrons bombard
interstellar nuclei of C, N, and O creating lighter isotopes.  It has been
suggested that the ``bullets'' and ``targets'' might be reversed near
supernovae where C, N, and O nuclei could be accelerated into the ambient
interstellar gas including protons and neutrons (see, for example, Duncan et
al.~1997, 1998, Lemoine, Vangioni-Flam \& Cass\'e 1998).  Modified and
expanded versions of Galactic cosmic ray spallation can be found, for
example, in Ramaty \& Lingenfelter (1999), Ramaty, Lingenfelter \& Kozlovsky
(2000), Parizot \& Drury (1999), Parizot (2000).

Although standard Big Band Nucleosynthesis (BBN) is not expected to produce
much Be (Be/H $\leq$10$^{-18}$), some models with inhomogeneous regions in
the very early universe can make up to Be/H $\sim$10$^{-14}$ (Malaney \&
Mathews 1992, Orita et al.~1997).  This amount is near the detection threshold
and efforts have been made to search for a plateau in Be similar to the Li
plateau (e.g. Boesgaard et al.~1999, Primas et al.~2000a).  The plateau in Li
abundance for metal-poor stars, from [Fe/H] = $-$1.5 to $-$3.5, has a small
dependence on metallicity and temperature found by Novicki (2005) in a sample
of 116 halo dwarfs.  A Be plateau might also show a similar dispersion.  If
there is a Be plateau, it need not be the result of production of Be in an
inhomogeneous Big Bang.  It could be caused by extra production of Be in
superbubbles by multiple supernova outbursts (Parizot \& Drury 1999).

The increase in Be with Fe may also show a dispersion in the Be abundance for
a given Fe.  Hints of this are seen in Figure 5 of Boesgaard et al.~(1999a),
particularly at [Fe/H] $\sim$ $-$1.5 where the spread in Be is $\sim$0.7 dex
but the typical errors are $\pm$0.10 dex.  A spread could result from
differing degrees of efficiency in the formation of Be by spallation in
different parts of the Galaxy.

We have made Be observations in a set of disk and halo stars to try to
understand the formation and Galactic evolution of Be.  Our sample includes
the very metal-poor dwarf, G 64-37, at [Fe/H] = $-$3.2 which will allow us to
examine whether there is a plateau in the Be abundances.  Other stars that we
observed enable us to investigate whether there is a spread in Be at given
metallicity.

\section{Observations and Data Analysis}

The Be II resonance lines at 3130.421 and 3131.065 \AA\ are best observed near
the meridian with a spectrometer and detector with good UV efficiency.  The
spectra for this research were obtained with the High Dispersion Spectrograph
(HDS) (Noguichi et al.~2002) on the Subaru 8.2 m telescope on Mauna Kea in 2001
June 05 and 2003 May 27.  The June 2001 night had less than optimal weather
(snow), so our observing program was limited to a total of nine stars over the
two nights.  The stars observed cover a broad range in metallicity: [Fe/H] =
+0.02 to $-$3.20.

The spectra were taken with the standard HDS setup StdUb with the blue
collimator and cross disperser, resulting in a wavelength coverage of
2970-4640 \AA\ over the two CCDs, with the echelle orders overlapping such
that the Be lines appeared in two orders in each exposure.  The slit width
chosen was 0.\arcsec7 wide, and the images were binned 2$\times$2 for
efficient readout.  This setup yields a spectral resolution of $\sim$50,000,
which is sufficient to measure the Be lines.  The log of the observations is
given in Table 1 along with photometric data for eight stars.  (The results
for the ninth star, G 186-26, have been published separately by Boesgaard \&
Novicki 2005.)

The majority of the data analysis for the Subaru HDS data was done using the
standard IRAF utilities.  For HDS, the overscan region is located in the
center of each chip, and the data for each CCD are read out using two
amplifiers.  The Subaru staff have written a routine for IRAF which removes
the overscan region and does a bias subtraction for each of the individual
images.  The master flat field was created by median-combining all of the
flats and then normalized using the the program {\em flatnormalize} in IRAF.
The flats that were taken on 2003 May 27 were saturated in the red region in
order to have sufficient flux in the region of the \ion{Be}{2} lines, and
therefore all the images taken that night were trimmed to avoid this region of
the CCD.  The saturated region did not contain the Be orders or those
immediately before or after, so by trimming the saturated region we did not
lose any of the Be data.  The June 2001 spectra were not saturated and
therefore all of the orders on the blue CCD were processed.

Once the flats were made and normalized, the images were divided by the master
flat.  The cosmic rays were removed and the profile of the scattered light
across the image was calculated and removed.  The spectra were traced and
extracted, along with the Th-Ar.  The list of Th-Ar lines in IRAF was
incomplete in the blue region of the spectra, so it was necessary to add many
lines to the list which we took from the Th-Ar atlas for HDS produced by the
Subaru telescope staff.  The wavelength solution from the Th-Ar spectra was
applied to the individual star spectra, and then the spectra were corrected
for their radial velocity shifts.  As the spectra are very crowded with
absorption lines, the radial velocities that were available via Simbad were
used to Doppler-correct the spectra.  Any multiple exposures for an individual
star were median combined, and then the spectra were divided by the continuum
level.  The Be lines are actually in two different orders of the spectrum, and
so if there was a cosmic ray event in the middle of the Be lines in one order,
the other order could be used.  For our analysis, we used the order that had
the Be lines near the center of the order, and therefore had more signal.  The
continuum level was estimated by fitting a high-order spline3 function to the
highest points in the spectrum because there were virtually no regions without
any absorption lines.  For two of the stars, G 21-22 and G 64-37, we combined
the two orders which contained the Be II lines in order to increase the
effective S/N.  Due to the strong blaze function, we had to combine the
``continuumed'' spectra weighted by the relative flux just shortward of the
blend containing the $\lambda$3130 line.

The reduced spectra covering the Be II lines are shown in Figures 1 and 2.  In
these figures the continua have been normalized to 1.0, but the final
continuum placement is done during the synthesis determinations.

\section{Abundances}

\subsection{Stellar Parameters}

For three of our stars we adopted published parameters.  For HD 195633 we used
the parameters of Rebolo et al.~(1988); for HR 7973 we used those from Chen et
al.~(2001); for HR 8888 we used those of Boesgaard et al.~(2001).  For the
other five stars we proceeded as follows.  We used three photometric colors as
temperature indicators: (b-y), (V-K), and (R-I).  The (b-y) colors came from
Schuster and Nissen (1988a, 1989a, 1989b) (hereafter SN), which was available
for all five stars.  The (V-K) and (R-I) colors came from a variety of sources
including the 2MASS survey (Cutri et al.~2003) and were also available for all
five stars.  Only one of our stars (G 21-22) needed a reddening correction,
which was derived by SN.  The required reddening correction for (b-y) was also
applied to the (V-K) and (R-I) using the relations from Johnson (1968).  We
have used the relationships between these colors and effective temperature
given by Carney (1983a) to find three values for T$_{\rm eff}$.  They were
then given weights of 4:2:1 for T(b-y)$_0$, T(V-K)$_0$, and T(R-I)$_0$.  These
temperatures are given in Table 2.  The error estimates come from averaging
the absolute difference between the temperature estimates and the mean T$_{\rm
eff}$.

The primary method used to estimate log g for the five stars was the use of
the star's position on the log g vs.~T$_{\rm eff}$ plane on a 10 Gyr, Y = 0.24
Isochrone.  We used the new Yale isochrones (Yi, Demarque \& Kim 2004).
First, the evolutionary status was determined from the Str\"omgren photometry
by placing the star on the c$_0$-(b-y)$_0$ diagram (see Schuster \& Nissen
1989a), and then the star was placed on the isochrone.  The isochrones are
double-valued for most temperatures and the evolutionary status indicated
which of the values was appropriate.  With the exception of HD 132475 the
stars are all dwarf stars.  

We estimate that the minimum error on log g is 0.20 dex, and we can be more
specific for some of the stars.  For HD 94028 Tomkin et al.~(1992) derived log
g from the ionization balance of Fe I and Fe II; their log g is 4.32.  Gratton
et al.~(1996) reevaluate the Tomkin et al.~parameters and find 4.54.
Fulbright (2000) determined stellar parameters spectroscopically and his
ionization balance value for log g is 4.2 for HD 94028.  Clementini et
al.~(1999) used an iterative procedure involving colors and Hipparcos data to
find a value for log g of 4.31 for this star.  Nissen et al.~(1997) found 4.30
$\pm$0.06 using a parallax-based gravity with Hipparcos data for HD 94028.
Our derived value of 4.44 $\pm$0.20 is consistent with these other
determinations; further, it is consistent with the way we have found log g
values for the other stars in our sample.

For HD 201891 we derive log g = 4.42 $\pm$0.20 from the isochrones.  Other
studies have found similar values: 4.31 from Clementini et al.~(1999); 4.43
$\pm$0.10 from Chen et al.~(2000); 4.30 $\pm$0.06 from Fulbright (2000), 4.25:
from Nissen et al.~(1997).  For HD 132475 we have found log g = 3.60
$\pm$0.20.  For this star Fulbright (2000) derived 3.60 $\pm$0.06 and Nissen
et al.~(1997) found 3.87 $\pm$0.10.  The other two stars, G 21-22 and G 64-37,
are too faint to have reliable parallaxes from Hipparcos.  For these two stars
we will adopt an uncertainty of a factor of two, i.e. $\pm$0.30 dex.
 
Once the temperatures were established for all of the sample stars, we
determined the [Fe/H] values.  We searched the literature for high-resolution,
high-S/N measurements of [Fe/H] in our five stars with lower metallicities.
In the Cayrel de Strobel et al.~(2001) catalog of [Fe/H] determinations from
high-resolution, high-S/N spectroscopic observations, for example, there are
13 entries for HD 94028, 7 for HD 132475, 15 for HD 201891, but only 2 for G
64-37 and 1 for G 21-22.  We have found additional references for the latter
two stars, including Alonso et al.~(1996) and Carney et al.~(1994) for G 21-22
and Stephens \& Boesgaard (2002) and Nissen et al.~(2003) for G 64-37.  All of
those results were normalized to solar Fe/H of 7.51 (on the log scale where
log N(H) = 12.00) and corrected to our temperature.  We have made calculations
with MOOG to determine the temperature sensitivity of [Fe/H] (e.g. see
Boesgaard et al.~(2005); it is typically $\pm$0.08 dex for $\pm$135 to 150 K
in T$_{\rm eff}$.  We adjusted the published [Fe/H] by the appropriate amount
to correspond to our derived temperature.  We also used some lower resolution
and/or lower S/N results, e.g. Th\'evenin \& Idiart (1999), Beers et
al.~(1999), but gave them less weight in our final averages.

For metal-poor stars near the turnoff point on the main sequence, the
microturbulent velocity, $\xi$, is nearly constant Magain (1984) and equal to
1.5 km s$^{-1}$, so we adopted this value for all of our models.  More details
on the parameter determinations can be found in Novicki (2005).

\subsection{Be Abundance Determinations}

Abundances were determined for Be through the spectrum synthesis mode of MOOG,
2002 version (Sneden 1973, http://verdi.as.utexas.edu/moog.html).  This
version contains the UV opacity edges of the metals from Kurucz.  There are
many atomic and molecular lines in the spectral region of the Be II lines.
The line list used by Boesgaard et al.~(1999a) was used here.  As pointed out
in that paper, it is important to know the O abundance as many lines of OH
appear among the blends and [O/Fe] is positive for metal-poor stars.
Therefore, for our syntheses the O abundance in the model atmosphere is not
reduced by the amount that Fe is reduced (the case for other elements, except
Be), but rather it is specified.  For stars where the O abundance has not been
specifically determined, the value has been found from the relationship
between [Fe/H] and [O/H] in Boesgaard et al.~(1999b).

We have determined A(Be) = log N(Be)/N(H) +12.00 for our stars.  The spectrum
synthesis fits for all eight stars are shown by pairs in Figures 3 - 6.  For
clarity, these figures contain a smaller wavelength interval than that which
we used to do the synthesis; the synthesis extends longward by another
\AA.  We adjusted four parameters (continuum level, wavelength shift,
broadening, and abundance) to get the best fit over this 3 \AA\ region.  To
find the Be abundance in the higher metallicity stars we relied on the longer
wavelength line, $\lambda$3131, because it it less blended than the
$\lambda$3130 line.  This can be seen in Figure 3 where the blended line is
not as well reproduced as the $\lambda$3131 line is.  (In addition, those two
stars have spectra that are broadened somewhat by rotation which increase the
blending effect.)  For the lower metallicity stars both lines could be used,
but the longer wavelength line was weighted more, except for G 64-37 with
[Fe/H] = $-$3.20 where all the lines are weak (including the Be II lines), so
we used the stronger, i.e.~shorter wavelength, of the two lines.  The adopted
Be abundances are given in Table 3 along with the errors (see next section).
Also included in Table 3 are the Li abundances for each of these stars taken
from Novicki (2005), Rebolo et al.~(1988), Chen et al.~(2001) and Boesgaard et
al.~(2001).

\subsection{Error Estimates}

We have estimated the uncertainty in the Be abundances from the errors of the
stellar parameter determinations and from the accuracy of the spectrum
synthesis fits.  To do this we have used the Kurucz grid models at
temperatures of 5750, 6000, 6250, and 6500 K, log g values of 4.0 and 4.5, and
a range in [Fe/H] values: +0.10, 0.00, $-$0.10, $-$0.20, $-$0.30, $-$0.50,
$-$1.00, $-$1.50, $-$2.00, $-$2.50, and $-$3.00.  We have tested the
dependency on A(Be) of the microturbulence parameter, $\xi$, by running models
with 2.0, 1.5, and 1.25 km s$^{-1}$.  Here is a brief summary: an error of
$\pm$80 K in T$_{\rm eff}$ gives an error in A(Be) of 0.01 (except there is no
change from 6000 - 6250 K at log g of 4.5); an error of $\pm$0.10 in [Fe/H]
gives an error of 0.03 - 0.04 in A(Be) (except at low [Fe/H] where a change of
0.5 in [Fe/H] gives an error of 0.01); and an uncertainty in log g of $\pm$0.2
produces errors in A(Be) from 0.07 (at the high temperature) to 0.10 (at the
lower temperatures in our models); a change in 0.25 km s$^{-1}$ in $\xi$ is
typically $\pm$0.00 - 0.02 in A(Be).  Thus the typical errors in A(Be) for our
stars are $\sigma_{T}$ $\sim$0.01, $\sigma_{Fe}$ $\sim$ 0.02, $\sigma_{\xi}$
$\sim$0.01, but $\sigma_{log g}$ $\sim$0.10.  The accuracy of the spectrum
synthesis fits is estimated from how large a change in A(Be) can still
represent the observed spectrum with adjustments in the continuum (which is
partly determined by the S/N of the spectrum) and gaussian smoothing in the
synthetic spectra.  This is typically 0.05 to 0.06 dex, but is 0.12 dex in G
64-37 which has S/N of only 87.  The final errors (added in quadrature) for
each star are given in Table 3.

In addition there may be uncertainties resulting from effects due to 1D vs.~3D
models as discussed by Primas et al.~(2000a).  According to the work of
Garc\'\i a L\'opez, Severino \& Gomez (1995) and Kiselman \& Carlsson (1995),
NLTE effects are found to be insignificant for Be.

\section{Results and Discussion}

The Be abundances determined here can be compared to those in the literature.
For HR 8888 we find A(Be) = +1.38 $\pm$0.10 in excellent agreement with 1.44
$\pm$0.13 in Boesgaard et al.~(2001) from CFHT/Gecko data.  For HD 94028 we
derive a value for A(Be) of +0.51 $\pm$ 0.12, agreeing with 0.54 $\pm$0.08
determined by Boesgaard et al.~(1999a) from Keck/HIRES data.  

Figure 7 plots A(Be) vs.~[Fe/H] for a larger sample of stars from several
sources to put our results in context.  In order to make meaningful
comparisons the data shown are all on the same temperature scale, use Kurucz
models, and use the same line list for the spectrum synthesis; they are
results from Boesgaard et al.~(1999a), Boesgaard (2000), Boesgaard et
al.~(2004).  Three of the lowest metallicity stars (shown as open squares in
the figure) are from Primas et al.~(200a, 200b), who also used Kurucz models
and our temperature scales.  Our stars are superposed on the plot as large
filled squares.  The five stars with [Fe/H] $>$ $-$1.10 fit well in the
context of the relationship between Be and Fe.  The large square in the upper
right is our value for HR 8888.  Our large square labeled HD 94028 is offset a
small amount from the circle below it from the Boesgaard et al.~(1999a) value.
Our result is a confirmation of the high Be abundance found earlier.  It also
indicates that the spread in A(Be) at this metallicity ([Fe/H] $\sim$ $-$1.5)
is real; Boesgaard et al.~(1999a) remarked in their discussion of this star
that its high Be abundance might indicate an intrinsic dispersion in Be at a
given Fe.  Two other stars, HD 132475 and the low metallicity star, G 64-37,
deserve individual discussion; they are shown with labels in Figure 7.

\subsection{HD 132475}

Novicki (2005) had observed that HD 132475 (with [Fe/H] = $-$1.50) has a Li
abundance, A(Li) = 2.39$\pm$0.04, that is more than 3$\sigma$ above the
``plateau'' abundance (2.24 dex at its metallicity and temperature).  As a
result, we chose to measure the Be abundance of this star to see if its Be
abundance is also greater than expected.  Figure 5b shows the fit to the Be
spectrum for this star, and it shows the best-fit abundance of A(Be) = 0.57.
When plotted against the data from the literature (as shown in Figure 7), this
star has a Be abundance which is about 0.5 dex above the other stars at that
metallicity, but similar to HD 94028.  There is another star in the Novicki
(2005) Li sample which has a Li abundance significantly greater than the
plateau value, BD +23 3912 at A(Li) = 2.51$\pm$0.01.  The abundances in this
star are discussed in detail by King, Deliyannis \& Boesgaard (1996).  They
measured abundances of other elements (C, O, Na, Al, Eu, Y, Zr, Ba, La, Nd,
Sm) and found them to be normal.  This star has also been observed for Be by
Boesgaard et al.~(1999a) and has normal Be.  These two stars (HD 132475 and BD
+23 3912 have similar atmospheric parameters --- T$_{\rm eff}$ = 5765 K,
[Fe/H] = $-$1.50, and log g = 3.60 for HD 132475, and T$_{\rm eff}$ = 5695,
[Fe/H] = $-$1.53, and log g = 3.77 for BD +23 3912 --- and therefore might be
expected to have similarly large Be abundances.  In fact, BD +23 3912 has a Be
abundance which is perfectly in line with the other stars at that metallicity:
A(Be) = 0.18.  It is possible that the Li enrichment is due to the
$\nu$-process in supernovae which produces Li and B, but not Be (Woosley et
al.~1990, Timmes et al.~1995).

One possible explanation for the high measured Li and Be abundances for HD
132475 is that the T$_{\rm eff}$ we use may be too high.  It is not likely,
though, that the T$_{\rm eff}$ has been dramatically overestimated; the
spectroscopic temperature (5765 K) and the photometric temperature (5777 K)
are in excellent agreement with each other.  Also, the literature estimates
for this star's temperatures range from 5550-5788 K, which is consistent on
the high end with our estimate (and, as seen in $\S$3.2, A(Be) is only weakly
dependent on temperature).  The other major factor in the Be abundance
estimate is log g, which we get from our spectroscopic analysis to be equal to
3.60.  Bonifacio \& Molaro (1997) find log g= 3.65, Hartmann \& Gehren (1988)
find log g = 3.50 and Fulbright (2002) finds 3.60.  These literature values
agree with our log g within our error of 0.2 dex.

HD 132475 was included in the sample of Fulbright (2000, 2002) who studied the
abundances and kinematics of a large group of field stars.  He found [Fe/H] =
$-$1.59 (similar to our $-$1.50), but A(Li) = 2.23 $\pm$0.10.  His lower Li
abundance results from his use of T$_{\rm eff}$ = 5575 K which is 190 K cooler
than ours and the higher temperature would bring it into alignment with our Li
value.  In comparison to his other stars at the metallicity of HD 132475, we
see that it is in the upper range of abundances, especially for [Na/Fe],
[Mg/Fe], [Si/Fe], [Y/Fe], [Zr/Fe], and [Ba/Fe].  His lower temperature only
affects the abundances of those elements by small amounts ($-$0.02 to +0.07
dex) according to his Table 8 (Fulbright 2000), but those abundances in HD
132475 are typically 0.2 dex above the mean for the other stars at the
metallicity and Galactic rest-frame velocity, v$_{rf}$ = 156 km s$^{-1}$ of HD
132475.  The enhancement of the alpha-elements, Mg and Si, indicates Galactic
chemical evolution (GCE) has occurred which would also result in an enrichment
of Li and Be.

The increase of Be with Fe shown in Figure 7 indicates that Be is enriched
over time by galaxy-wide processes.  However, stars like HD 132475 and HD
94028 seem to indicate that there may be additional local enrichments, perhaps
by spallation in the vicinity of Type II supernovae where C, N, and O atoms
may be accelerated into the local gas and are broken into smaller atoms such
as Li, Be, and B.  Given the typical errors bar shown in 
Figure 7, there does seem to be a spread in A(Be) at a given [Fe/H].  The idea
that the CNO atoms can act as the ``bullets'' in spallation reactions was
suggested by Duncan et al.~(1997) in their paper on B abundances in the
Galaxy.

In a series of papers by Parizot (2000, 2001) and Parizot \& Drury (1999,
2000) these authors promote the idea that the light elements can be formed
efficiently inside a superbubble with multiple supernovae.  In this case the
accelerated CNO atoms bombard the ambient material to create the ``pieces''
which are Li, Be, and B atoms (along with other products like $^{11}$C, He
isotopes, etc.).  In the 2000 paper Parizot \& Drury discuss a bimodal
production of the light elements: ``standard'' GCR production (originally
suggested by Reeves, Fowler \& Hoyle 1970) and spallation reactions from a
collection of supernovae inside a superbubble.  They point to the similar pair
of stars, HR 94028 and HD 219617, from Boesgaard et al.~(1999a) which differ
in A(Be) by 0.6 dex although they have similar temperatures, metallicities and
log g values; the ``excess'' Be in HD 94028 could be the ``local''
contribution from the superbubble component.  We have confirmed the excess Be
abundance in HD 94028 and now add HD 132475 as another candidate for extra
enrichment from a superbubble of supernovae.

It is hypothesized by Fields et al.~(2002) that hypernovae could produce large
amounts of the light elements, Li, Be, and B from their exploding C-O cores.
This mechanism provides the local enrichment of the light elements, including
Be, in the next generation of stars.  In these hypernovae the composition is
basically C and O, the parent nuclei of the light elements.  Furthermore, the
high kinetic enery released would accelerate the envelope sufficiently for
spallation to occur with C and O being the ``bullets.''  Hypernovae candidates
have been observed, e.g. by Iwamoto et al.~(1998, 2000), Kawabata, et
al.~(2003) and Mazzali et al.~(2002, 2004) and may be associated with
gamma-ray bursts.

\subsection{G 64-37}

A number of authors (Thorburn 1994, Ryan et al.~1996, 1999) have observed that
G 64-12 and G 64-37 are two dwarf stars with several characteristics in
common.  They have similar temperatures (their temperatures differ from 20 K
to 200 K depending on the author, with G 64-12 always being hotter),
metallicities ($-$3.6 vs. $-$3.2), log g values and locations on the sky.
They have one key difference -- their Li abundances are different by more than
a factor of two: 2.40 vs. 2.06 (Novicki 2005).

With no obvious explanation for the difference in the Li abundances, we turn
to Be to see if there are any clues there.  Primas et al.~(2000a) measured the
Be abundance in G 64-12 to be A(Be) = $-$1.15 $\pm$0.15, which is above the
mean trend of Be vs.~[Fe/H] at that metallicity (shown in Figure 7).  These
authors argue that the Be abundances of the stars at this metallicity
strengthen the idea of a possible flattening in the Be-Fe trend -- a possible
plateau.  A plateau in Be abundance at low metallicity could indicate that, as
an analog to the Li plateau, some Be was produced in an inhomogeneous BBN.  A
high Be abundance could also result from some enrichment process as discussed
in the previous section (e.g. Parizot \& Drury 2000, Fields et al.~(2002).

Novicki (2005) finds the Li abundance in G 64-12 to be 2.40 $\pm$0.03 which is
+0.3 dex above the Li plateau.  If the abundance of Li in G 64-12 has been
augmented to lift it significantly above the Li plateau value, it is also
possible that the Be abundance could have been augmented as well by the same
process.  This star, G 64-12, is key to the argument that there is a Be
plateau at low metallicities.  If it does have augmented Li and Be, by
definition the measured abundances of these elements are not primordial, and
it should not be used in any plateau analysis.  For G 64-37 Novicki (2005)
derives a Li abundance of +2.06 $\pm$0.04, or less than half the amount in
G64-12.

We determine the abundance of Be in G 64-37 to be A(Be) = $-$1.30 $\pm$0.19,
which is lower than the abundance measured by Primas et al.~(2000a) for G
64-12 of A(Be) = $-$1.15 $\pm$0.15, but not different within the errors.  The
spectrum for this star is shown in Figure 6, along with four syntheses at
$-$0.70, $-$1.00, $-$1.30, and $-$$\infty$.  The stronger Be II line at 3130
\AA\ is clearly present and $-$1.30 seems to give a reasonable fit.  A higher
S/N spectrum may allow us to make a more accurate determination.  So both Li
and Be are more abundant in G 64-12 than in G 64-37.  The three stars at very
low metallicity which have measurable Be abundance are G 64-12 as measured by
Primas et al.~(2000a), CD $-$24 17504 by Primas et al.~(2000b) and this work
on G 64-37.

It has been suggested by Vangioni-Flam et al.~(1998) that the lowest
metallicity stars may have increased Li and Be resulting from shock
acceleration of light nuclei due to multiple supernovae in a superbubble.  The
excesses above the plateau for these oldest stars would come from the
explosions of early generations of very massive stars, 60 -- 100 $M_{\odot}$.
Since only G 64-12 is enhanced and apparently not G 64-37, the superbubble
giving rise to G 64-12 apparently did not include G 64-37.

\section{Summary and Conclusions}

We have determined Be abundances in eight stars from high-resolution, high-S/N
spectra from the 8.2 m Subaru telescope using HDS.  The stars cover the
metallicity range [Fe/H] from +0.02 to $-$3.20.  The most metal-rich star, HR
8888, has the meteoritic abundance of Be, i.e.~it is undepleted.  The four
stars with [Fe/H] between $-$0.3 and $-$1.1 fall perfectly along the
previously established relationship between A(Be) and [Fe/H].  There are two
stars, HD 94028 and HD 132475, with [Fe/H] = $-$1.5 and these both fall
significantly above the observed trend of A(Be) with [Fe/H].  With HD 94028 we
confirm the previous observation of a high Be abundance from Keck/HIRES
observations (Boesgaard et al.~1999a).  HD 132475 was observed for Be because
we had found its Li abundance to be 3$\sigma$ higher than other Li-plateau
stars at its metallicity.  The Be abundance of HD 132475 is 4$\sigma$ above
the Fe-Be trend.  (Another star, BD +23 3912, which has similar stellar
parameters and a high Li abundance has normal Be, not the same high Be
abundance as HD 94028 and HD 132475.)  Now there are two stars with Be
abundances higher than typical at [Fe/H] = $-$1.5 by some 0.5 - 0.6 dex; this
implies that there is a cosmic dispersion in the Be abundances, probably as a
result of the environment around the star at the time of its formation.

The general trend in Be vs.~Fe shows that there are Galaxy-wide processes at
work that increase Be and Fe over time, but the two stars above the trend show
that there are local enrichments in addition.  Such Be enhancements could be
caused by extra Be production by spallation in superbubbles from multiple
supernova explosions (Parizot \& Drury 2000) or in the vicinity of hypernovae
(Fields et al.~2002).

There is some evidence in the most metal-poor stars that there is a plateau in
the Be abundance.  We have added a third star to the ``plateau investigation''
below [Fe/H] = $-$3.0 with our observation of G 63-37 with [Fe/H] = $-$3.20.
This star has A(Be) = $-$1.30 which puts it in line with the general Fe-Be
trend, and {\it not} evidence of a Be plateau.

We compared the two stars G 64-12 and G 64-37, which have similar temperatures
and gravities and very low metallicities ([Fe/H] $\sim$ $-$3.3).  They have
different light element abundance patterns.  G 64-12 has higher Li than the
plateau value at A(Li) = 2.40, while G 64-37 is lower than the Li plateau at
A(Li) = 2.06.  Primas (2000a) found a high Be abundance in G 64-12 of A(Be) =
$-$1.15, above the trend, while our A(Be) for G 64-37 of $-$1.30 is consistent
with the general trend.  This adds to the evidence for a dispersion in Be at a
given metallicity.

We suggest that there are two types of spallation.  One is responsible for
the general increase of Be with Fe (or O) due to Galaxy-wide spallation.  The
other results from local enhancements by spallation near multiple supernovae
or hypernovae.

\acknowledgments 

We thank the Subaru support staff for their enthusiastic help before, during
and after the observations.  This work has been supported by NSF grants
AST-0097945 and AST-0505899 to AMB.

\clearpage

\singlespace
\begin{center}
\begin{deluxetable}{lrrrcc} 
\tablewidth{0pc}
\tablecolumns{7} 
\tablecaption{Subaru HDS Observations} 
\tablehead{ 
\colhead{Star}  &  \colhead{V}  & \colhead{B-V}  & 
\colhead{Night} & \colhead{Exp. Time} & \colhead{Total}  \\

\colhead{} & \colhead{} & \colhead{}  & \colhead{} & \colhead{(min)} & \colhead{S/N}} 
\startdata 
HD 94028    &   8.23   & 0.471 &  27 May 2003  & \phn 10 & 134 \\
G 64-37     &  11.14   & 0.370 &  27 May 2003  & 270 & \phn 87 \\
HD 132475   &   8.57   & 0.577 &  27 May 2003  & \phn 20 & \phn 96 \\
G 21-22     &  10.72   & 0.527 &  27 May 2003  & \phn 90 & \phn 96 \\
HD 195633   &   8.54   & 0.523 &  06 Jun 2001  & \phn\phn  5 & 143 \\
HR 7973     &   6.01   & 0.430 &  06 Jun 2001  & \phn 25 & 146 \\
HD 201891   &   7.38   & 0.510 &  06 Jun 2001  & \phn 14 & 188 \\
HR 8888     &   6.61   & 0.360 &  27 May 2003  & \phn\phn 3 & 116 \\
\enddata 

\end{deluxetable} 
\end{center}


\clearpage

\singlespace
\begin{center}
\begin{deluxetable}{lccccccc} 
\tablewidth{0pc} 
\tablenum{2} 
\tablecolumns{8} 
\tablecaption{Stellar Parameters}
\tablehead{ \colhead{Star} & \colhead{T$_{\rm eff}$} & \colhead{$\sigma$(T)} 
& \colhead{$\log{\rm g}$}  & \colhead{$\sigma$(log g)} &
\colhead{[Fe/H]} & \colhead{$\sigma$([Fe/H])} & \colhead{$\xi$} \\ 

\colhead{} & \colhead{(K)} & \colhead{(K)} & \colhead{(dex)} &
\colhead{(dex)}  &\colhead{(dex)} & \colhead{(dex)}  
& \colhead{(km~s$^{-1}$)} 
}
\startdata 
HD 94028  &  5907 & $\pm$69   & 4.44 & $\pm$0.20 & $-$1.54 & $\pm$0.10 & 1.5 \\
G 64-37   &  6233 & $\pm$69   & 4.41 & $\pm$0.30 & $-$3.20 & $\pm$0.23 & 1.5 \\
HD 132475 &  5765 & $\pm$62   & 3.60 & $\pm$0.20 & $-$1.50 & $\pm$0.15 & 1.4 \\
G 21-22   &  5916 & $\pm$84   & 4.59 & $\pm$0.30 & $-$1.02 & $\pm$0.25 & 1.5 \\
HD 195633 &  5986 & $\pm$115  & 3.89 & $\pm$0.08 & $-$0.88 & $\pm$0.16 & 2.0 \\
HR 7973   &  6339 & $\pm$70   & 4.20 & $\pm$0.10 & $-$0.31 & $\pm$0.10 & 1.8 \\
HD 201891 &  5806 & $\pm$110  & 4.42 & $\pm$0.20 & $-$1.07 & $\pm$0.10 & 1.2 \\
HR 8888   &  6722 & $\pm$50   & 4.00 & $\pm$0.20 & +0.02   & $\pm$0.10 & 2.5 \\
\enddata 
\end{deluxetable} 
\end{center}


\clearpage

\singlespace
\begin{center}
\begin{deluxetable}{lcccccc} 
\tablewidth{0pc} 
\tablenum{3} 
\tablecolumns{7} 
\tablecaption{Abundances}
\tablehead{ \colhead{Star} & \colhead{T$_{\rm eff}$} & 
\colhead{[Fe/H]} & \colhead{A(Li)}  & \colhead{Li
ref\tablenotemark{a}}  & \colhead{A(Be)} & \colhead{$\sigma$(Be)} \\ 

\colhead{} & \colhead{(K)} & \colhead{(dex)} &
\colhead{(dex)}  & \colhead{}  & \colhead{dex}  
& \colhead{dex}
}
\startdata  
HD 94028    &  5907  & $-$1.54  & 2.25 & BN  & 0.51 & 0.12 \\
G 64-37     &  6233  & $-$3.20  & 2.06 & BN & $-$1.30 & 0.19 \\
HD 132475   &  5765  & $-$1.50  & 2.39 & BN  & 0.57 & 0.12 \\
G 21-22     &  5916  & $-$1.02  & 2.48 & BN  & 0.33 & 0.16 \\
HD 195633   &  5986  & $-$0.88  & 2.15 & RMB & 0.66 & 0.11 \\
HR 7973     &  6339  & $-$0.31  & 2.17 & C01 & 1.02 & 0.10 \\
HD 201891   &  5806  & $-$1.07  & 1.98 & BN  & 0.62 & 0.11 \\
HR 8888     &  6722  & +0.02    & 3.30 & B01 & 1.42 & 0.10 \\

\enddata 

\tablenotetext{a}{BN = this work; RMB = Rebolo et al.~(1988); C01 = Chen et
al.~(2001); B01 = Boesgaard et al.~(2001)}

\end{deluxetable} 
\end{center}


\clearpage

\begin{figure}
\plotone{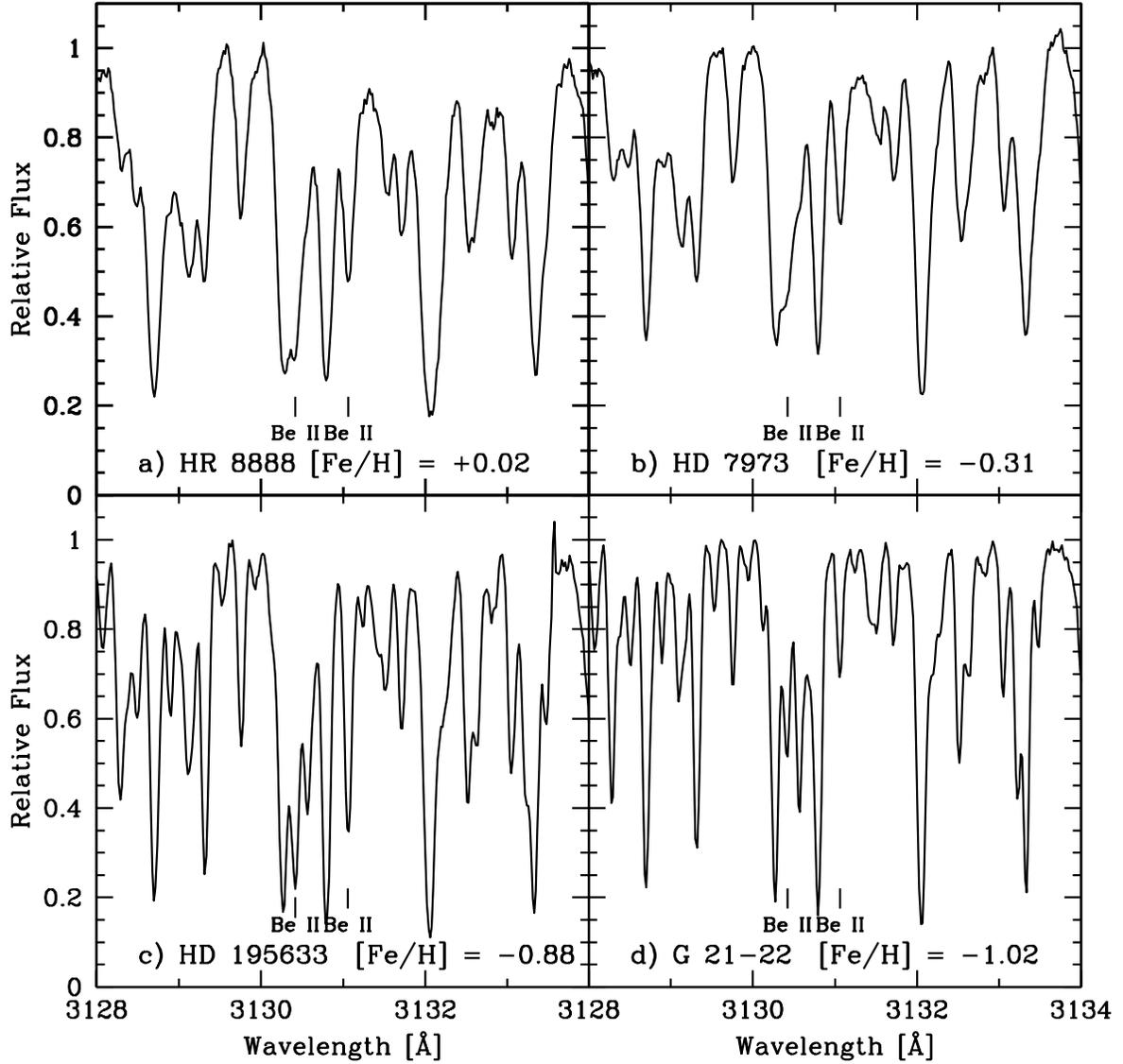}
\caption{Spectra of the region around the Be II lines in four of the higher
metallicity stars.  The spectra in panels a) and b) are broadened somewhat by
rotation.  The spectra have been normalized to a continuum of 1.0, but this is
adjusted in the synthesis fits.}
\end{figure}

\begin{figure}
\plotone{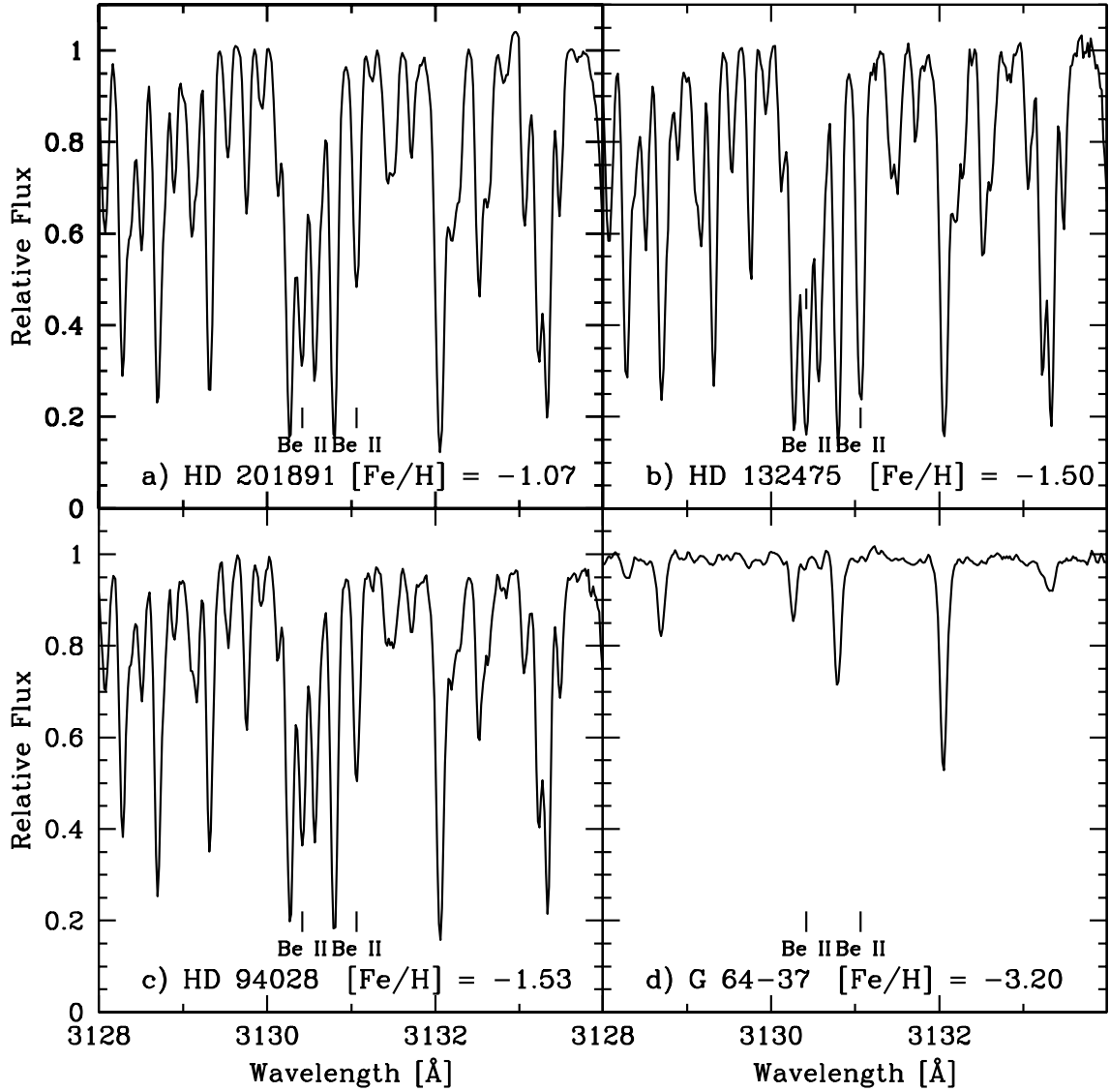}
\caption{Spectra of the region around the Be II lines in four of the lower
metallicity stars.  The spectra have been normalized to a continuum of 1.0,
but this is adjusted in the synthesis fits.}
\end{figure}

\begin{figure}
\plotone{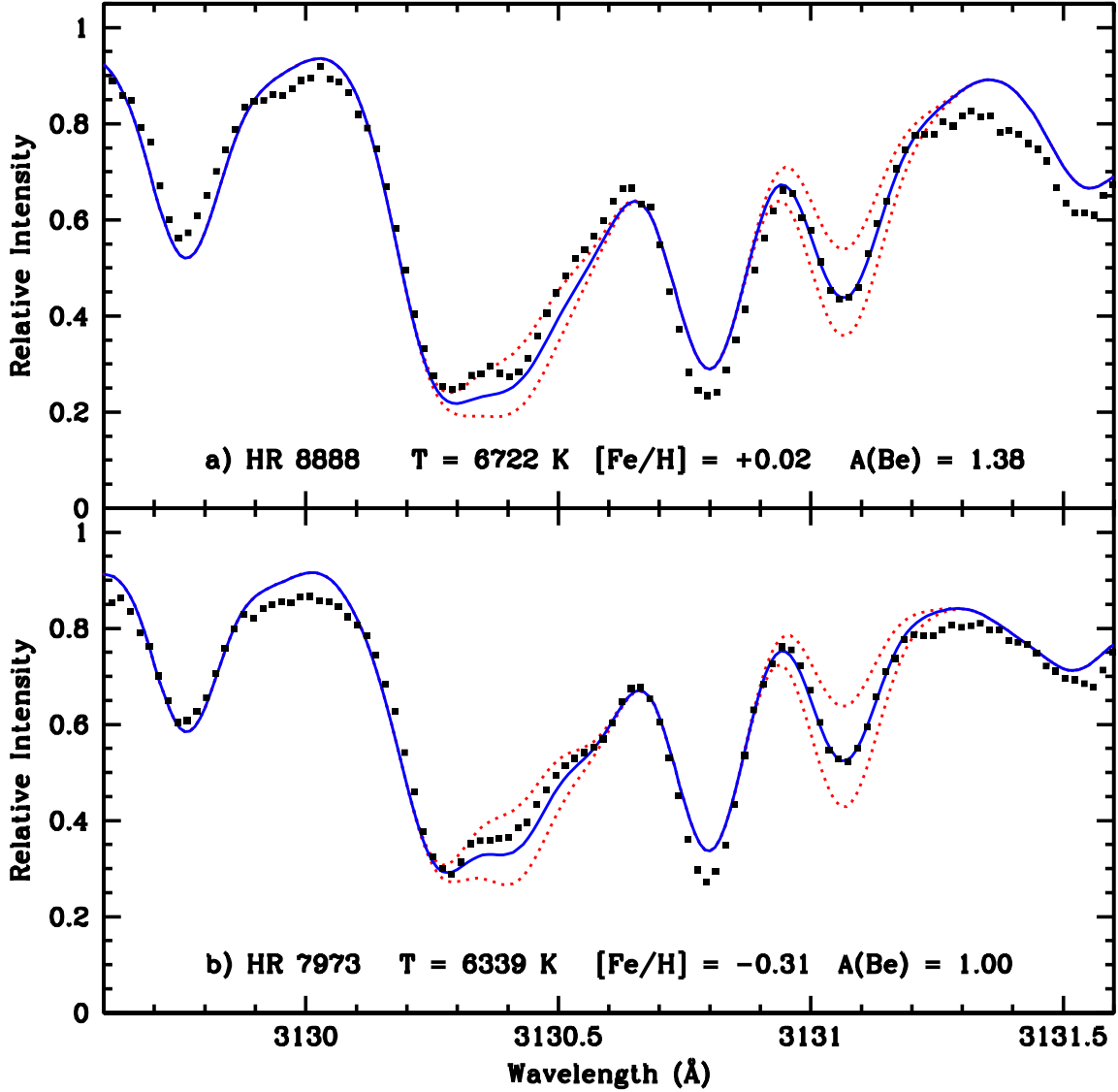}
\caption{Spectrum synthesis fits for the two highest metallicity stars.
The spectra of these two stars are broadened by rotation.  The small filled
squares are the data points, the solid line is the best fit synthesis and the
dotted lines are a factor of 2 higher and lower values for A(Be).  Because the
blending in the shorter wavelength line is so severe, we relied on the longer
wavelength line to find the Be abundance.}
\end{figure}

\begin{figure}
\plotone{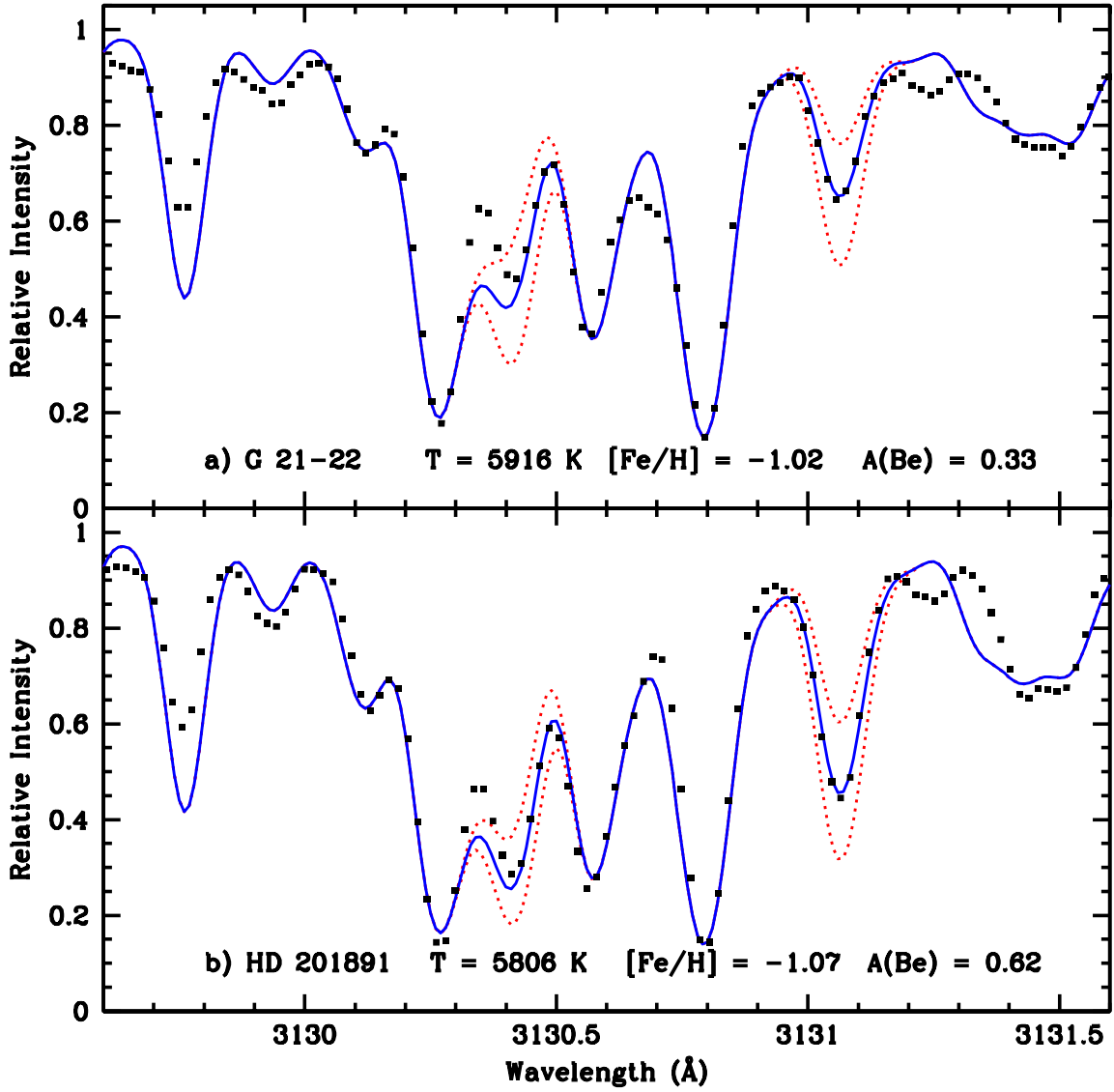}
\caption{Spectrum synthesis fits for two similar intermediate-metallicity
stars.  The symbols and lines are the same as in Figure 3.  Again we relied on
the less-blended, longer-wavelength Be line.  Although these stars are similar
in temperature and metallicity, they differ in their Be abundances by a factor
of 2.}
\end{figure}

\begin{figure}
\plotone{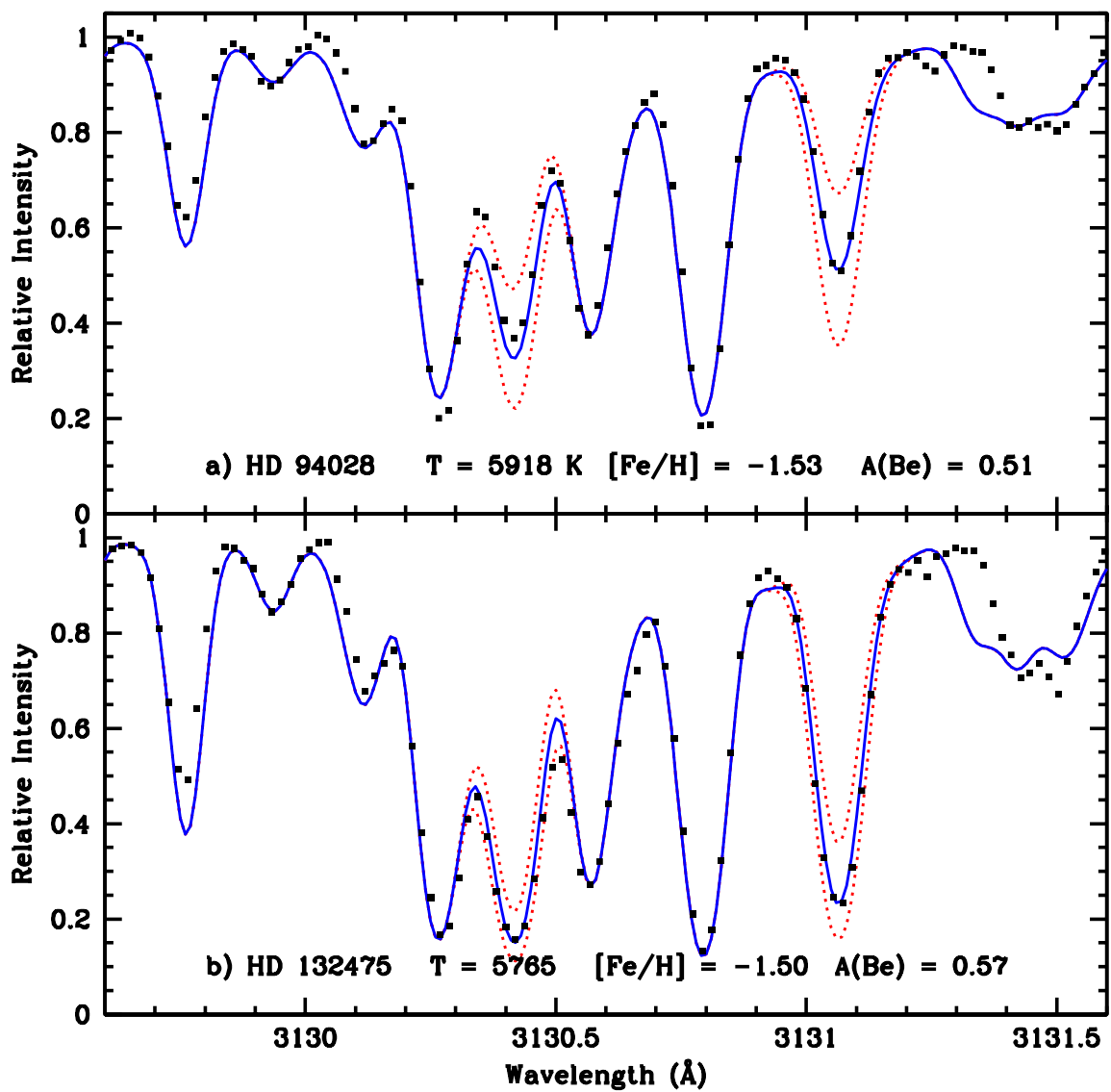}
\caption{Spectrum synthesis fits for two lower metallicity stars with similar
values of [Fe/H].  The symbols and lines are the same as in Figure 3.  Like
the stars in Figure 4, these two stars are also similar in temperature and
metallicity, but their Be abundances are essentially the same.  The Be II
lines strengths differ, but that is due to the effect of log g: For HD 132475
log g = 3.60, while for HD 94028 log g = 4.44.}
\end{figure}

\begin{figure}
\plotone{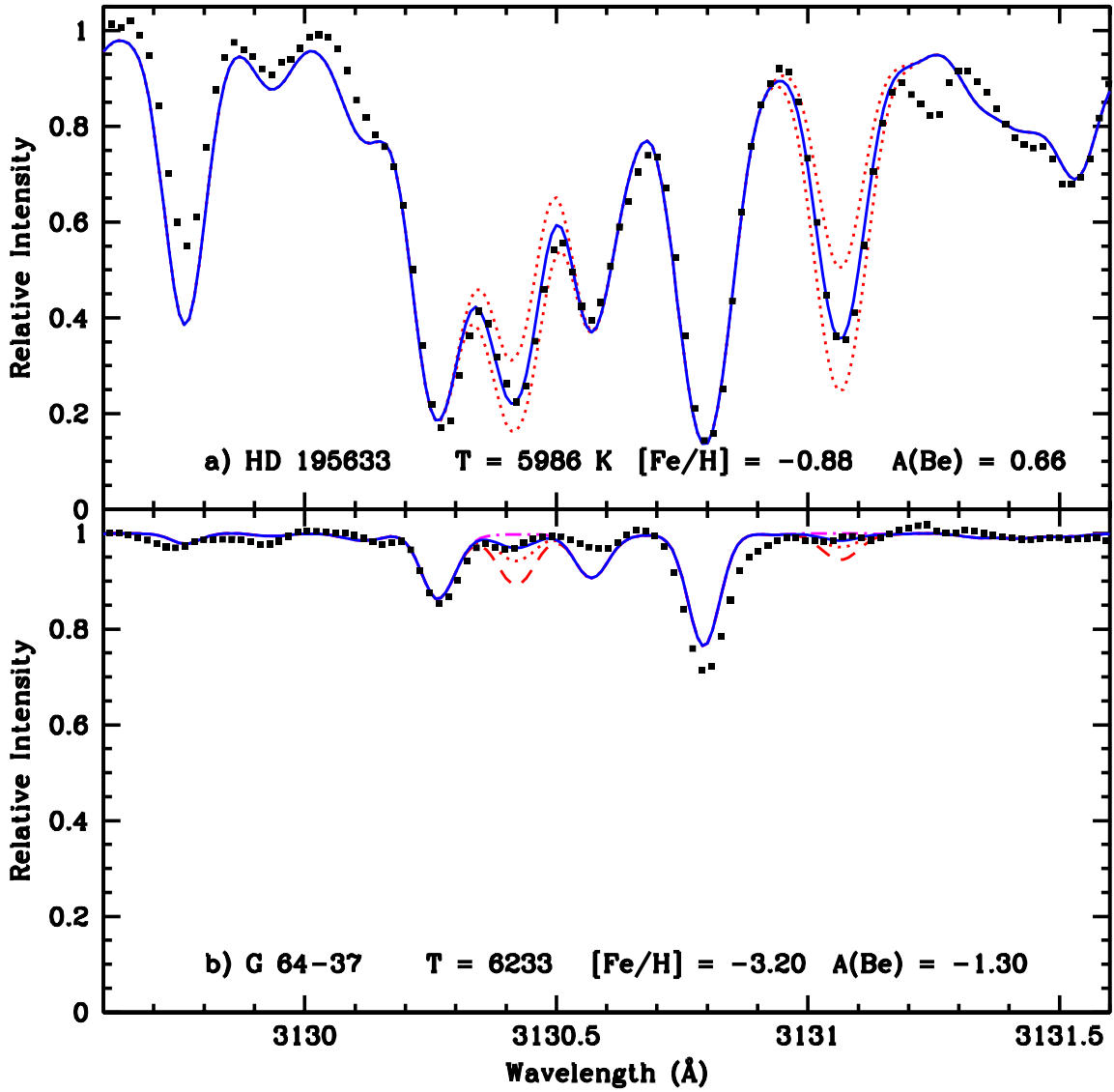} 
\caption{Spectrum synthesis fits for two stars with very different
metallicities.  For panel a) the symbols and lines are the same as in Figure
3.  But for G 64-37 we show four synthesis fits at A(Be) = $-$0.70, $-$1.00,
$-$1.30, and $-$$\infty$.  The best fit (the solid line) is at $-$1.30.  In
this case, we relied of the stronger line (the shorter wavelength line) to
find the Be abundance. }
\end{figure}

\begin{figure}
\plotone{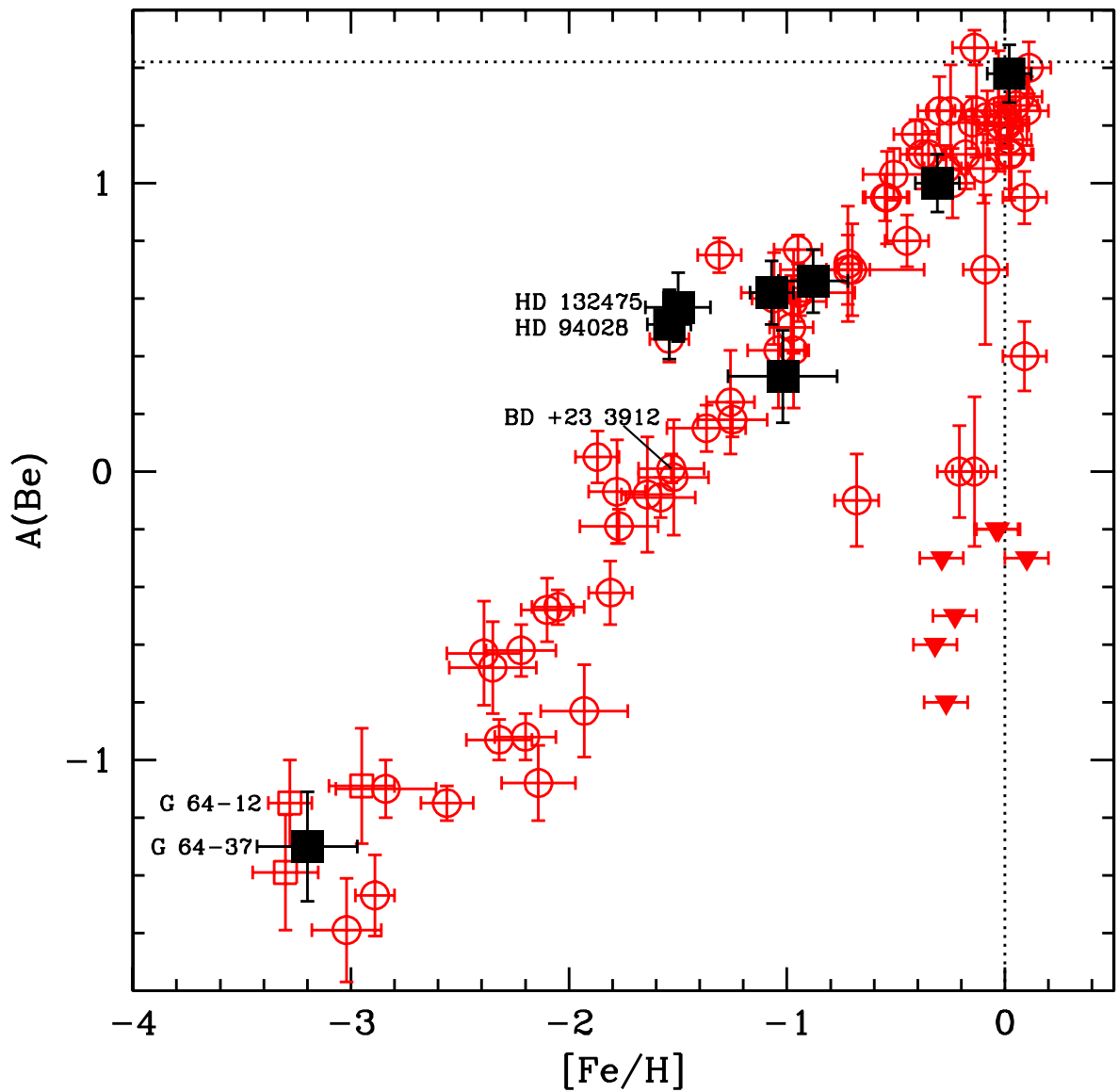}
\caption{A(Be) vs. [Fe/H] for this sample compared to the literature sample.
Our results are the large filled squares superposed on this diagram.  Open
circles and filled triangles (upper limits) are from Boesgaard et al.~(1999a),
Boesgaard (2000), Boesgaard et al.~(2004).  Open squares (low metallicity
stars) are from Primas et al.~(2000a, 2000b).  Individual error bars on both
A(Be) and [Fe/H] are shown.  The horizontal dotted line is the meteoritic Be
abundance, A(Be) = 1.42 (Grevesse \& Sauval 1999).}
\end{figure}

\end{document}